\documentclass{elsart}
\usepackage{amssymb}
\usepackage{bbm}
\usepackage{psfrag}
\usepackage{graphicx}
\usepackage{latexsym}
\newcommand{\be}{\begin{eqnarray}}
\newcommand{\ee}{\end{eqnarray}}

\newcommand{\Tr}{\rm Tr}
\newcommand{\const}{\rm const}
\usepackage{amsfonts}
\usepackage{epsf}
\usepackage{amsmath}

\begin{document}

\begin{frontmatter}

\title{Signal and Noise in Financial Correlation Matrices}

\author{Zdzis\l aw Burda},
\ead{burda@th.if.uj.edu.pl}
\author{Jerzy Jurkiewicz\thanksref{now}}
\ead{jurkiewicz@th.if.uj.edu.pl}
\thanks[now]{Presented by J.J. at {\em Applications of Physics in Financial 
Analysis 4}
Conference, Warsaw, 13-15 November 2003}

\address{M. Smoluchowski Institute of Physics, Jagiellonian University, 
Reymonta 4, 30-059~Krak\'ow, Poland.}

\begin{abstract}

Using Random Matrix Theory one can derive exact relations 
between the eigenvalue spectrum of the covariance matrix and
the eigenvalue spectrum of it's estimator (experimentally measured
correlation matrix). These relations will be used to analyze a 
particular case of the correlations in financial series and to
show that contrary to earlier claims, correlations can be measured
also in the ``random'' part of the spectrum. Implications for the
portfolio optimization are briefly discussed.

\end{abstract}

\end{frontmatter}

\section{Introduction}
\label{intro}

Empirically determined correlation matrices appear in many research areas.
In financial analysis they became one of the cornerstones of the
financial risk analysis through the idea of the optimal portfolio by
Markowitz \cite{mar}. The practical way to obtain the correlation
matrix is to perform a (large) number $T$  of measurements of a (large)
number $N$ of experimental quantities (price fluctuations) 
$x_{it},~i=1,\dots N,~t=1,\dots T$ and to
estimate the correlation matrix by
\be
c_{ij} = \frac{1}{T} \sum_{it} x_{it}x_{jt},
\label{def}
\ee
where we assumed that $\langle x_i\rangle = 0$.
The relation of the estimated correlation matrix $c_{ij}$ to the true
correlation matrix $C_{ij}$, in particular the relation of the eigenvalue
spectra of the two matrices was addressed in the literature (cf. e.g. 
\cite{MP,SM,SB,BGJJ}), assuming the experimentally
measured quantities come from some correlated random process.
We shall address this point in the next section, assuming these distributions
come from the correlated Gaussian ensemble.

\begin{figure*}[htp]
\begin{center}
\includegraphics[width=10cm]{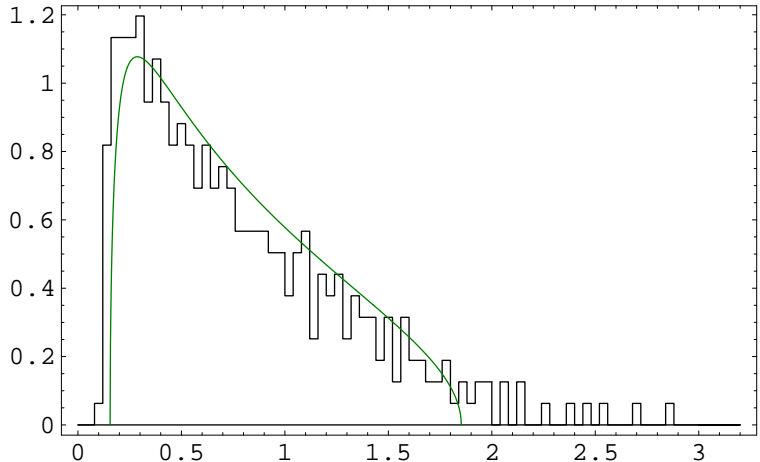}
\caption[Fig.1]{\label{fig1}
Eigenvalue distribution of the normalized correlation matrix following 
\cite{LCBP}. Nine largest eigenvalues are missing. The line represents a RMM
distribution with a single eigenvalue.}
\end{center}
\end{figure*}

In the context of financial analysis, the correlation matrices were discussed
in \cite{GBP,LCBP,PGRAS,PK,LM,BJN}.
The general conclusion, based on results of \cite{LCBP} was rather pessimistic.
Let us quote here from the paper \cite{PK}: {\em \dots covariance matrices
determined from empirical financial time series appear to contain such a high
amount of noise that their structure can essentially be regarded as random.
This seems, however, to be in contradiction with the fundamental role played 
by covariance matrices in finance, which constitute the pillars of modern
investment theory and have also gained industry-wide applications in risk
management}. Let us recall the arguments which led to this conclusion.
In the paper \cite{LCBP} the authors analyzed the S\&P500 price fluctuations 
of $N=406$ assets in the period of $T=1309$ days between 1991-1996. Price
fluctuations were normalized to the standard deviation $\sigma_i=1$ and the
correlation matrix was constructed, following (\ref{def}). The spectrum
of the $406 \times 406$ correlation matrix contains a number of large
eigenvalues, the largest of which was interpreted as a ``market''. The
spectrum, where the nine largest eigenvalues are missing is presented
on figure \ref{fig1}.
The lower part of the spectrum was interpreted as a distribution of eigenvalues
of a random uncorrelated Gaussian matrix, therefore lacking any information 
about the correlations.
In the following we will try to dissolve this pessimism. In the next section
we will present a simple model of correlated Gaussian fluctuations and
discuss the properties of the resolvent of this model. We will use the
Random Matrix Theory solution \cite{BGJJ} to construct explicitly the
eigenvalue density function $\rho(\lambda)$ in case the spectrum of
the correlation matrix is known. We shall also discuss the problem
of obtaining this spectrum from the experimentally measured eigenvalue
distribution.
We will apply this method to
reanalyze the financial time series discussed above.
Summary and conclusions will be given in the last section.

\section{The correlated Gaussian fluctuations.}
\label{model}

Let us consider a Gaussian model, generating a matrix $X$ of fluctuations
$x_{it},~i=1,\dots N,~t=1,\dots T$ with a matrix measure
\be
P(X) DX = \frac{1}{Z}\exp(-\frac{1}{2}\Tr X^T C^{-1}X)
\label{measure}
\ee
where $Z$ is a normalization factor and $C$ is a real symmetric positive
correlation matrix, representing the genuine two-point correlations in
the system. A single matrix $X$ generated from this ensemble can be used
to estimate the spectrum of $C$ by (\ref{def}). Matrix $C$ can always be
diagonalized and we assume the set of  it's eigenvalues $\{\mu_i\}$ to be 
known. We shall define two resolvents $G(Z)$ and $g(z)$ defined by
\be
G(Z)= \frac{1}{N}\Tr(Z-C)^{-1},~~~g(z) =\frac{1}{N}\langle\Tr(z-c)^{-1}\rangle,
\label{resolv}
\ee
where the averaging is made with the measure (\ref{measure}). In (\ref{resolv})
$Z$ and $z$ are complex variables. In the limit $N\to\infty,~T\to\infty~
N/T=r < 1,~r=\const$ we can use the diagrammatic technique \cite{BGJJ}
to find the {\bf exact} relation between the two resolvents. To take
the limit for $G(Z)$ we assume that eigenvalues appear in blocs with degeneracy
$k_j = p_j N,~j=1,\dots M$ and the number of blocs $M$ stays finite. Function
$G(Z)$ is an analytic function of $Z$ on a complex plane. It has poles at
$Z=\mu_j$ with  residues $p_j$. Similarly $g(z)$ is an analytic function on 
the complex $z$ plane with one or more cuts along the real axis. 
Discontinuities across these cuts are related to the eigenvalue density
$\rho(\lambda)$ by
\be
\rho(\lambda)=\frac{1}{\pi}{\rm Im}g(\lambda+i0^+).
\ee
The exact relation between the resolvents has a form of a duality
relation
\be
z g(z)=ZG(Z),
\label{dual}
\ee
where the complex arguments $z$ and $Z$ are related by the conformal map
\be
Z=z(1-r+r z G(z)),~~z=\frac{Z}{1-r+r Z G(Z)}.
\label{map}
\ee
On the ``physical'' sheet on the $z$ plane $z \to \infty$ for $Z \to \infty$
and as follows from (\ref{map}) the poles at $Z=\mu_j$ are mapped to $z=\infty$
on other Riemann sheets of $g(z)$. 

The relation between $g(z)$ and $G(Z)$ can be used to find the eigenvalue
distribution $\rho(\lambda)$ {\bf for any set} $\{\mu_j,p_j\}$. 
Using (\ref{map}) we find the position of the
cuts on the $z$ plane by relating them to the map of the ${\rm Im} z=0^{\pm}$
lines on the $Z$ plane. We then find the imaginary part of the resolvent $g(z)$
using (\ref{dual}).  


It is more difficult to obtain  $\{\mu_j,p_j\}$ from the measure distribution
$\rho(\lambda)$. The eigenvalue distribution is known only approximately, 
from one realization of the estimator $c_{ij}$. It is nevertheless possible
to extract more information about the spectrum of the correlation matrix,
than suggested by the pessimistic approach quoted above. Both resolvents
contain information about the moments of the correlation matrices. Expanding
around $Z=\infty$ (resp. $z=\infty$) we have
\be
G(Z)=\sum_{i=0} \frac{1}{Z^{i+1}}\frac{1}{N}\Tr C^i,~~~
g(z)=\sum_{i=0} \frac{1}{z^{i+1}}\frac{1}{N}\langle\Tr c^i\rangle.
\ee
For $r<1$ we have similar expansions around $Z=0$ (resp. $z=0$). Duality 
relation can be viewed as relation between the moments
\be
M_i = \sum_{j=1}^M p_j\mu_j^i~~{\rm and }
~~m_i=\int d\lambda\rho(\lambda)\lambda^i.
\ee
For the first few moments we have:
\be
M_1 &=& m_1 \nonumber \\
M_2 &=&  m_2 -r m_1^2 \nonumber \\
M_3 &=& m_3-3rm_1 m_2 +2 r^2 m_1^3\\
\cdots \nonumber
\ee
and
\be
M_{-1} &=& (1-r) m_{-1} \nonumber \\
M_{-2} &=& (1-r)^2 m_{-2}-r(1-r)m_{-1}^2 \nonumber \\ 
M_{-3} &=& (1-r)^3 m_{-3}-r(1-r)^2m_{-1}m_{-2} -r^2(1-r)m_{-1}^2\\
\cdots \nonumber
\ee
These relations can be used to determine the moments of the correlation
matrix $C$ from the moments of it's estimator $c$. It is clear that high 
moments will have larger errors.

As an example let us reconsider the analysis of the S\&P500 data presented
in the Introduction. In the analysis we assumed six eigenvalue blocs in
the correlation matrix, which 
gave the optimal fit of the experimental distribution.
\begin{figure*}[htp]
\begin{center}
\includegraphics[width=10cm]{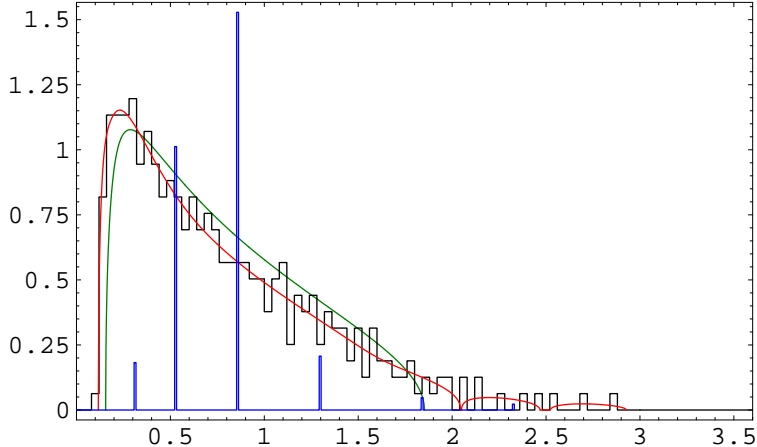}
\caption[Fig.2]{\label{fig2}
Analysis of the eigenvalue distribution presented above. We assumed six
eigenvalues, which gave the optimal fit of the distribution. The determined
eigenvalues are represented by blue lines and their relative height
corresponds to the probability $p_j$. The red line is the fit of the 
distribution with six eigenvalues, to be compared with the green line, 
corresponding to the fit presented before.}
\end{center}
\end{figure*}

\section{Discussion and summary}

Experimental estimator of the correlation matrix, although containing
statistical noise, can be nevertheless used to extract much more information
about the spectrum of the true correlation matrix than is usually believed.
This information can be very important when the optimized portfolio is
constructed. In case the matrix $C$ is diagonal, one can derive exact
relations between the volatility $\sigma_0$ of the true optimal portfolio, 
obtained from the correlation matrix $C$, volatility $\sigma_R$ of the
portfolio based on the estimate $c$ and the volatility $\sigma$ predicted
from this estimate. This relation depends on $r$
\be
\sigma_R = \frac{\sigma_0}{\sqrt{1-r}} = \frac{\sigma}{1-r}
\ee
and shows that, particularly for $r$ close to 1, the error can be quite
substantial.

If the spectrum of the correlation matrix $C$ is known, we can always predict
the shape of the measured eigenvalue distribution $\rho(\lambda)$ of it's
estimator $c$. Although the formulas presented above are valid in principle
only for infinitely large matrices, in practice the predicted spectra agree 
very well even for the matrix sizes of the order 100 and below. 

It is also possible to invert the problem and to determine the the spectrum
of $C$ from that of the estimator $c$, although this step can be done only 
approximately. 

\vspace{1cm}
{\bf Acknowledgement.} 

This work was partially supported by the EC IHP Grant
No. HPRN-CT-1999-000161, by the Polish State Committee for
Scientific Research (KBN) grants 
2P03B 09622 (2002-2004) and 2P03B 08225 (2003-2006),
and by EU IST Center of Excellence "COPIRA".




\begin{thebibliography}{00}
\bibitem{mar} H. Markowitz, {\em Portfolio Selection: Efficient Diversification
of Investments}, Wiley, New York, 1959.
\bibitem {MP} V. A. Marcenko and L. A. Pastur, Math. USSR-Sb, (1967) 457.
\bibitem{SM} J. Silverstein, J. Multivariate Anal. {\bf 30} (1989) 1.
\bibitem{SB} J. Silverstein and Z. D. Bai, J. Multivariate Anal {\bf 54} (1995)
175.
\bibitem{GBP} G. Gallucio, J,-P, Bouchaud and M. Potters, Physica {\bf A259}
(1998) 449.
\bibitem{LCBP} L. Laloux, P. Cizeaux, J.-P. Bouchaud and M. Potters, Phys. Rev.
Lett. {\bf 83} (1999) 1467.
\bibitem{PGRAS} V. Plerou, P. Gopikrishnan, B. Rosenow, L. A. N. Amaral and
 H. E. Stanley, Phys. Rev. Lett. {\bf 83} (1999) 1471
\bibitem{PK} S. Pafka and I. Kondor, Eur. Phys. J. {\bf B27} (2002) 277,
cond-mat/0205119.
\bibitem{LM} F. Lilo and R. N. Mantegna, cond-mat/0305546.
\bibitem{BJN} Z. Burda, J. Jurkiewicz and M. A. Nowak, Acta Phys. Polon.
{\bf B34} (2003) 87.
\bibitem{BGJJ} Z. Burda, A. G\"{o}rlich, J. Jarosz and J. Jurkiewicz,
cond-mat/0305672.







\end{thebibliography}
\end{document}